\documentclass[oldversion]{aa} 
\usepackage{graphicx}
\usepackage{lscape}
\usepackage{verbatim}
\usepackage{multirow}
\usepackage{lscape}
\usepackage{txfonts}
\usepackage{booktabs}
\usepackage{bigstrut}
\usepackage{natbib}
\usepackage{indentfirst}
\bibpunct{(}{)}{;}{a}{}{,}

\newcommand{\tten}[1]{\times 10^{{#1}}}
\newcommand{\msun}{$M_{\odot}$}

\newcommand{\lsun}{$L_{\odot}$}

\newcommand{\tff}{$\tau_\mathrm{ff}$}
\newcommand{\hh}{H$_2$}
\newcommand{\brg}{Br$\gamma$}
\newcommand{\uflux}{erg~s$^{-1}$~cm$^{-2}$}

\begin{document}
   \title{From large scale gas compression to cluster formation in the Antennae overlap region \thanks{Based on observations with the VLT/SINFONI, Program ID 383.B-0789, with the VLT/CRIRES, Program ID386.B-0942, and with the CFHT/WIRCAM, Program ID 09AF98}}

   \author{C. N. Herrera,
          F. Boulanger
          \and
          N. P. H. Nesvadba
          } 
   \institute{Institut d'Astrophysique Spatiale, UMR 8617 CNRS,
Universit\'e Paris-Sud 11, 91405 Cedex Orsay, France}

\abstract
{We present a detailed observational analysis of how merger-driven
turbulence may regulate the star-formation efficiency during galaxy
interactions and set the initial conditions for the formation of super
star clusters. Using VLT/SINFONI, we obtained near-infrared imaging
spectroscopy of a small region in the Antennae overlap region,
coincident with the supergiant molecular cloud 2 (SGMC~2). We find
extended \hh\ line emission across much of the 600 pc field-of-view,
traced at sub-arcsecond spatial resolution.  The data also reveal a
compact \hh\ source with broad lines and a dynamical mass
$M_\mathrm{dyn}\sim 10^7~M_{\odot}$, which has no observable \brg\ or
$K$-band continuum emission, and no obvious counterpart in the 6~cm
radio continuum.  Line ratios indicate that the H$_2$ emission of both
sources is powered by shocks, making these lines a quantitative tracer
of the dissipation of turbulent kinetic energy. The turbulence appears
to be driven by the large-scale gas dynamics, and not by feedback from
star formation. We propose a scenario where the H$_2$ emission is
related to the formation of bound clouds through accretion. The
kinetic energy of the accreted gas drives the turbulence and powers
the H$_2$ emission. Broad H$_2$ line widths of-order 150 km s$^{-1}$,
similar to the velocity gradient of the gas across the field of view,
support this idea. Within this interpretation, the compact \hh\ source
could be a massive cloud on its way to form a super star cluster
within the next few Myr. This scenario can be further tested with ALMA
observations.}

\authorrunning{Herrera, C. N.  et al.}
\maketitle
%
%


\section{Introduction}

Detailed observations of the nearby Antennae galaxy merger
NGC4038/4039 at a distance of only 22 Mpc \citep{schweizer08} provide
important benchmarks for our understanding of how the large-scale
dynamics of galaxy interactions trigger star formation on much
smaller scales. Contrary to the 'canonical' scenario of merger-induced
star formation, which predicts intense starbursts in the nuclear
regions \citep[e.g.,][]{barnes96}, most stars in the
Antennae are formed off-nucleus, in a heavily obscured, gas-rich,
infrared-luminous region where both galaxies permeate each other, the
``overlap region''\citep[e.g.,][]{vigroux96,klaas10}.

Most of the star formation in the overlap region of 20~\msun~yr$^{-1}$
\citep{zhang01} occurs in massive super-star clusters (SSCs) with
masses larger than $10^4$ and up to a few $10^6 \, $\msun\
\citep{whitmore95,zhang99,whitmore10}. Detailed studies of SSCs
constrain the recent star formation history of the overlap region. At
typical ages of few $10^6$ yrs, many SSCs have already expelled most
of the gas and dust from their immediate surroundings, and they may be
becoming unbound themselves \citep{fall05,mengel05,gilbert07,fall10}.

Our understanding of how the galaxy interaction may trigger gas
collapse and the very early phases in the formation of SSCs, is much
less developed and relies mostly on theoretical arguments. For
example, \citet{scoville86} suggested that the collision between the
two galaxies could trigger collisions between pre-existing giant
molecular clouds. \citet{jog88} argued that shock-heated ambient gas
may enhance star formation by increasing the pressure in embedded
molecular clouds. \citet{keto05} discuss their high resolution
interferometric observations of molecular gas and the formation of
massive star clusters in the starburst galaxy M82 within this
scenario.  However, observations of the Antennae overlap region may
not be accounted for by the collapse of pre-shock clouds. The large
surface density of molecular gas and its fragmentation in complexes
with masses of several $10^8\,$\msun\ \citep{wilson00}, two orders of
magnitude larger than masses of giant molecular clouds in spiral
galaxies, are evidence for cooling and gravitational fragmentation of
the diffuse gas compressed in the galaxy collision.

First attempts to quantify the impact of the interaction on gas
cooling and fragmentation have been made with numerical simulations
\citep{teyssier10, karl10b,karl10}, but their description of the
energy dissipation of the post-shock, turbulent multiphase ISM is
still schematic. The simulations follow the turbulent energy of the
gas over a limited range of scales.  The mechanical energy of the
galaxy collision is thermalized in large-scale shocks and radiated
away by the shock compressed post-shock gas.  Observations of the
galaxy-wide shock in Stephan's Quintet \citep{appleton06,cluver10}
show that a dominant fraction of the collision energy is not
thermalized in such shocks, but cascades to smaller scales into
turbulent motion within molecular gas.  This results from the
clumpy multi-phase structure of the pre-shock interstellar medium
\citep{guillard09}.  The fact that most of the collision energy is
transferred to the molecular gas has consequences for the energy
dissipation and the gravitational fragmentation of the post-shock gas,
and thereby for induced star formation, which have yet to be
understood.  It is the motivation of this paper to investigate how gas
compression and turbulence driven by the large-scale dynamics of the
two interacting galaxies trigger and regulate star formation in the
Antennae overlap region.

Bright H$_2$ line emission powered by shocks (rather than star
formation) has been identified in the diffuse interstellar medium of
the Milky Way \citep{falgarone05}, as well as in a significant number
of extragalactic systems including many interacting galaxies
\citep{appleton06, cluver10, zakamska10}.  This makes emission-line
observations of warm H$_2$ an interesting tracer of the dissipation of
turbulent energy in the molecular gas.  \citet{haas05} proposed, based
on ISO observations, that H$_2$ line emission in the overlap region
could probe large-scale shocks driven into molecular clouds. This was
later called into question by \citet{brandl09}, who found lower H$_2$
line fluxes with Spitzer-IRS, and suggested that the bulk of the warm
molecular gas may be heated by star formation after all. However, at a
spatial resolution of $\sim$ 5\arcsec, and a spectral resolution of
${\rm R=600}$ ($\sim500$ km s$^{-1}$) Spitzer-IRS does not allow to
study the molecular gas at the scales relevant for the fragmentation
of the molecular gas, $\le$100 pc and $\sim$100 km s$^{-1}$,
respectively.

We use near-infrared spectroscopy obtained with the imaging
spectrograph SINFONI and the echelle spectrograph CRIRES (both on the
ESO Very Large Telescope) to quantify the dissipation of kinetic
energy in the molecular gas on scales of few 10s to 100s of pc. To
this end we observed the region around ``knot 5'' of \citet{brandl09},
the brightest H$_2$ peak in the overlap region. This knot, which is
near the massive and IR-bright cluster No. 28 of \citet{whitmore10},
coincides with the super-giant molecular cloud SGMC~2 observed by
\citet{wilson00} in CO. SGMC~2 is one of the most massive molecular
clouds in the overlap region, with a mass of $4\tten{8}\, $\msun. Our
pointing also coincides with knot K2a in the Herschel dust imaging of
\citet{klaas10}. Thus, overall, by selecting an area with particularly
bright line emission of warm molecular hydrogen, we observed one of
the main sites of molecular gas and star formation in the Antennae
overlap region.

The paper is organized as follows. The observations and data reduction
are described in Sect. 2. In Sect. 3 we focus on the structure and kinematics
of the warm \hh\ gas.  We identify two components of \hh\ emission:
diffuse emission extended over the 600~pc wide field of view of
SINFONI, and a compact source. We
show that, for the compact and diffuse extended emission, the \hh\
emission is powered by shocks rather than UV heating, and estimate the
bolometric \hh\ luminosity of each component combining Spitzer and
SINFONI observations (Sect. 4). In Sect. 5 we propose an interpretation of the
observations which relates the H$_2$ emission to the formation of bound
clouds through gas accretion. In Sect. 6 we discuss the observations and
our interpretation within a broader astrophysical context,
highlighting the role of turbulence for the formation of super star
clusters and the regulation of the star formation efficiency in galaxy
mergers. Section~7 gathers the main conclusions.


\begin{figure*}[ht] \centering
\includegraphics[width=11.9cm]{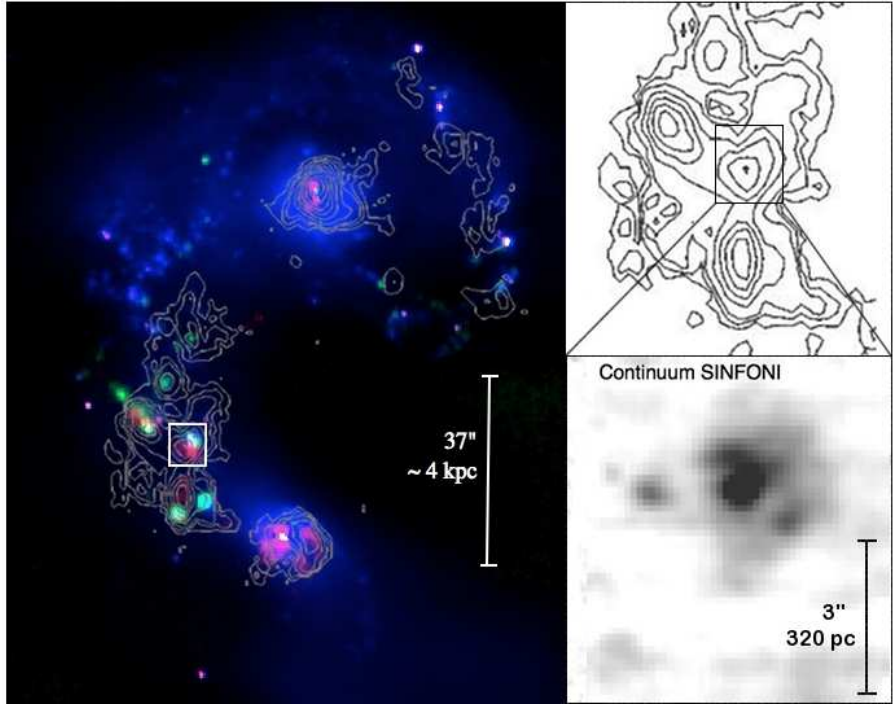}
\caption{Left panel: Central regions of the Antennae galaxy merger
including both nuclei. The K-band continuum, H$_2$ 1$-$0 S(1), and
Br$\gamma$ line emission are shown in blue, red, and green,
respectively.  All data were taken with WIRCAM on the CFHT. Grey
contours show the CO(1$-$0) line emission from \citet{wilson00}. The
white square represents our SINFONI pointing. The H$_2$ 1$-$0 S(1)
emission has the same morphology as the Spitzer IR emission in
\citet[][their Fig.~4]{brandl09}. The H$_{2}$ morphology does not
follow the continuum morphology, however, the Br$\gamma$ 
does. Right panel: Zoom onto the CO map of the overlap region (top,
the square shows our SINFONI field of view) and the line-free $K$-band
continuum morphology seen with SINFONI (bottom). North is up, and east
to the left in all panels.}\label{fig:pos}
\end{figure*}

\section{Observations}

\subsection{VLT/SINFONI imaging spectroscopy}

Our analysis is based on observations carried out with the
near-infrared integral-field spectrograph SINFONI
\citep{eisenhauer03,bonnet04} on the ESO Very Large Telescope. Data
were obtained in service mode under Program ID 383.B-0789 (PI
Nesvadba). SINFONI is an image slicer with a field of view of
8\arcsec$\times$8\arcsec\ and a pixel scale of 250~mas in the
seeing-limited mode. The spectral resolution is ${\rm R=4000}$ at
$\lambda=2.2~\mu$m. We observed three regions in the Antennae
galaxy merger in the $K$-band, one near each nucleus and one region in
the overlap region. These pointings were selected based on their
bright pure-rotational H$_2$ line emission detected with
Spitzer/IRS. This paper presents an analysis of molecular emission of
the pointing in the overlap region.

All data were taken in June and July 2009 under good and stable
atmospheric conditions.  We obtained a total of 1800 seconds of
exposure time split into individual exposures of 600 seconds. We
adopted a dither pattern where one sky frame was taken in-between two
object frames to allow for an accurate subtraction of the night
sky. Data reduction was done with the standard IRAF tools to reduce
longslit spectroscopy \citep{tody93}, which we extended with a set of
SINFONI-specific IDL routines. For details see, e.g.,
\citet{nesvadba07, nesvadba08}. The telluric correction was done using
bright ($K\sim6-8$ mag) stars observed at a similar air mass as our
target. We used the light profiles of these stars to measure the PSF
of our data, ${\rm FWHM=0\farcs7\times0\farcs6}$. Comparison with
high-resolution HST/NICMOS imaging of \citet{whitmore10} shows that
our PSF estimate is robust.

\subsection{CRIRES high-resolution spectroscopy and CFHT narrow-band imaging}

To complement our SINFONI observations we also obtained a
high-resolution longslit spectrum with CRIRES on the VLT (program ID
386.B-0942; PI Herrera) and near-infrared imaging of the Antennae
through the H$_2$ 1$-$0 S(1) and the \brg\ filters with WIRCAM at the
Canada-France Hawaii Telescope (CFHT program ID 09AF98; PI Nesvadba).

With CRIRES we obtained a total of 5~hrs of integration time of the
H$_2$ 1$-$0 S(1) line at 2.12~$\mu$m in the rest-frame for one
pointing centered on the compact molecular source (Sect. ~\ref{sec:cs})
and at position angle ${\rm PA=67^{\circ}}$. Atmospheric conditions
were good and stable. We nodded our target along the slit to ensure a
robust sky subtraction. Data were dark-subtracted, flat-fielded, and
sky subtracted. The wavelength calibration was done on telluric OH
lines straddling the wavelength of the H$_2$ 1$-$0 S(1) line. Using a
0.4\arcsec\ wide slit, we reached a spectral resolution of~6 km
s$^{-1}$, a factor $\sim$20 higher than the effective resolution of
our SINFONI data. This allowed us to analyze the line profile of the
compact molecular source at a spectral resolution which is comparable
to that reached by \citet{wilson00} for CO.

We determined the astrometry of the near-IR CFHT images through
cross-correlation with the 2MASS $K$-band image of NGC4038/39. We
give a short description of these data in Sect. \ref{sec:ec}. They have
also been used to improve the astrometric accuracy of the SINFONI
observations (Sect. \ref{sec:ali}).

\subsection{Ancillary data sets}\label{archive}

We complement our near-IR spectro-imaging data with CO(1$-$0)
observations obtained with the Caltech Millimeter Array. C. Wilson
generously shared her zeroth and first-moment maps of NGC4038/4039
with us. For a full discussion of these observations, see
\citet{wilson00, wilson03}. Our SINFONI pointing in the overlap region
coincides with their super-giant molecular cloud 2 (SGMC 2). Within
the spatial resolution of the CO data, the peak of SGMC~2 coincides
with the warm H$_2$ peak \#5 in \citet{brandl09}. In our analysis we
adopt the mass, size and line width of SGMC~2 listed in Table~1 of
\citet{wilson00}.

In addition, we use published results from Spitzer/IRS mid-infrared
spectroscopy by \citet{brandl09}. Our pointing corresponds to their
Peak \#5. In Sect. \ref{sec:lum} we will use the rotational H$_{2}$ line
fluxes from their Table~5 to compute the bolometric H$_{2}$
luminosity. The integrated S(2) flux measured with the Short-High (SH)
module is a factor 1.4 higher than that measured with the Short-Low
(SL) module.  \citet{brandl09} attribute this difference to the
different apertures used for the SH and SL flux measurements.  Since
the aperture of the SH module ($4\farcs4 \times 11\farcs3$) is best
matched to the SINFONI field of view, we scale the S(3) flux, which
was measured only with the SL module, by a factor 1.4.

\subsection{Astrometry of the SINFONI images}\label{sec:ali}
The SINFONI field-of-view is too small to register these data directly
with astrometric catalogues. Therefore, to ensure that the absolute
positional accuracy of our data is better than 1\arcsec, we collapsed
the SINFONI cube over the wavelength range of our CFHT $Ks$-band
image, which we previously registered relative to the USNO catalog of
astrometric standards, and matched the coordinates of both images.
This gives a positional accuracy of $\sim$0\farcs5 for the SINFONI
data. The position of the super-star cluster, the brightest continuum
source in the SINFONI field of view, is within 0\farcs5 of archival
HST/NICMOS images (taken through the F187N and F160W filters).
Curiously, we find a 2\arcsec\ offset relative to the coordinates of
the super star cluster given by \citet[][cluster \#28 in their
Table~8]{whitmore10}

 
\begin{table*}
\caption{Parameters of the super star cluster in our field.}\label{tab:sc}
\begin{center}
\begin{tabular}{c c c c c c c c}
\hline \hline
\noalign{\smallskip}
Obs & Size       & Age & $A_{K}$ &Mass & $N_\mathrm{ion}$& K & Reference \\
         & \arcsec\ & Myr  & mag      & $\tten{6}$ \msun\ & $\tten{52}$ s$^{-1}$& mag& \\
\hline
\noalign{\smallskip}
HST  & $-$ & 4.8$^{a}$ & 0.8$^{a,b}$  & 1.2$^{a}$ & $-$ & $-$ &\#28 from Table 8 in [1]\\ 
Keck + VLT & 0\farcs46$^{c}$ & 5.7$^{d}$ & $-$ & 4.1$^{e}$ & 1.3$^{e}$ & 15.2$^{f}$ &SSC-C in [2] \\
\multirow{2}{*}{VLT/SINFONI}    & \multirow{2}{*}{0\farcs6$^{g}$} & $<$ 6  & \multirow{2}{*}{0.4} &\multirow{2}{*}{6.0} & 4.4$^{e,h}$  & \multirow{2}{*}{15.0$^{f}$} & \multirow{2}{*}{Our work [3]} \\
                                    &  & 4.5  &            &                     & 6.6$^{i}$  &        & \\
\hline
\end{tabular}
\end{center}

[1] \citet{whitmore10} based on ACS and NICMOS observations, [2]
\citet{gilbert07} based on Keck/NIRSPEC-SCAM and VLT/ISAAC
observations, [3] For details see Sect. \ref{sec:ec}.

$^{a}$ from the comparison of integrated photometry UBVIH$\alpha$ with
population synthesis models, $^{b}$ estimated from measured
$E(B-V)=2.20$ and assuming $R_{V}=3.1$ extinction curve
\citep{weingartner01}, $^{c}$ deconvolved FWHM from Lorentzian fit,
$^{d}$ based on the \brg\ equivalent width as an age indicator
comparing it with Starburst99 \citep{leitherer10} models, $^{e}$ not
corrected for extinction, $^{f}$ $K$-band apparent magnitude for an
aperture of 2\arcsec, $^{g}$ It is unresolved. The angular size (FWHM)
measured from a Gaussian fit corresponds to the size of the seeing
disk, $^{h}$ from observed \brg\ luminosity, $^{i}$ from the fluxes of
the [\ion{Ne}{ii}] and [\ion{Ne}{iii}] lines at 12.81 and
$15.56\,\mu$m .

\end{table*}

\section{Observational results}
In Sect. ~\ref{sec:ec}, we present the morphology and kinematics of
the molecular and ionized gas.  In the two following sections, we
highlight the presence of extended H$_{2}$ emission throughout
the field (with broad FWHM$\sim$150-200 km s$^{-1}$,
Sect. \ref{sec:ee}), and the discovery of a compact molecular source, with
bright \hh\ line emission, which shows no \brg, nor continuum emission
in the K-band (Sect. \ref{sec:cs}).

\subsection{Identification of individual components}\label{sec:ec}

\begin{figure}
\centering
\includegraphics[width=8cm]{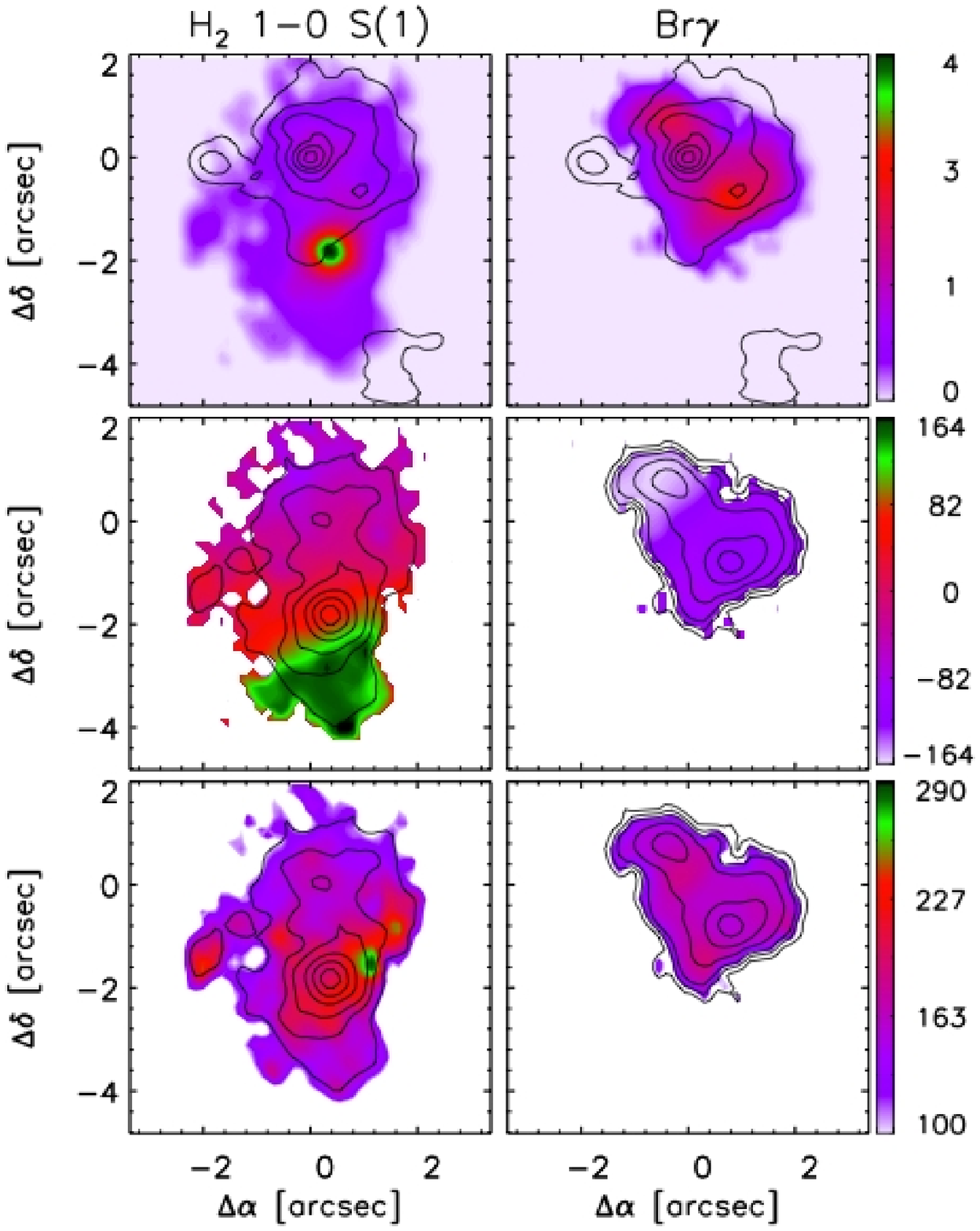}
\caption{H$_{2}$~1$-$0~S(1) and Br$\gamma$ morphologies and
kinematics. North is up, and East to the right in all panels. Top:
Line surface brightnesses in units of
$\times10^{-17}$~erg~s$^{-1}$~cm$^{-2}$.  The contours show the K-band
continuum in steps of 15\%, starting at 10\% of the peak intensity,
$45\tten{-20}$ erg s$^{-1}$ cm$^{-2}$ $\AA^{-1}$. Center: Velocities
in km~s$^{-1}$ relative to the mean recession velocity in our
field-of-view of 1556~km~s$^{-1}$. Bottom: Measured FWHMs in
km~s$^{-1}$. All maps are spatially smoothed by averaging the original
data over 3~pixels~$\times~$3~pixels (0$\farcs4\times0\farcs$4~
pix$^{-1}$). The contours in the mid and lower images show the line
surface brightness. Levels are 0.1, 0.2, 0.3, 0.5, 0.7, and 0.9 the
peak intensity (3.8$\times10^{-17}$~erg~s$^{-1}$~cm$^{-2}$ for \hh\
and 2.6$\times10^{-17}$~erg~s$^{-1}$~cm$^{-2}$ for \brg). Offsets are
relative to the peak in the $K$-band continuum map,
$\alpha(J2000): 12^{h}01^{m}54\fs753$, $\delta (J2000):
-18\degr52\arcmin51\farcs44$.
 }\label{fig:flx}
\end{figure}

The left panel of Fig.~\ref{fig:pos} shows the overall morphology of the
continuum-subtracted \hh~1$-$0 S(1) and \brg\ line emission in the
Antennae. \hh\ emission is found in the two nuclei and in the overlap
region. \brg\ is found in the overlap region with a different
morphology than the molecular gas. Our SINFONI observations targeted
the brightest peak in pure-rotational molecular line emission by
\citet{brandl09}. The field-of-view of our observations is marked by
the white box. The right panel of Fig.~\ref{fig:pos} shows two zooms
into this region.

\begin{figure}
\centering
\includegraphics[width=6cm]{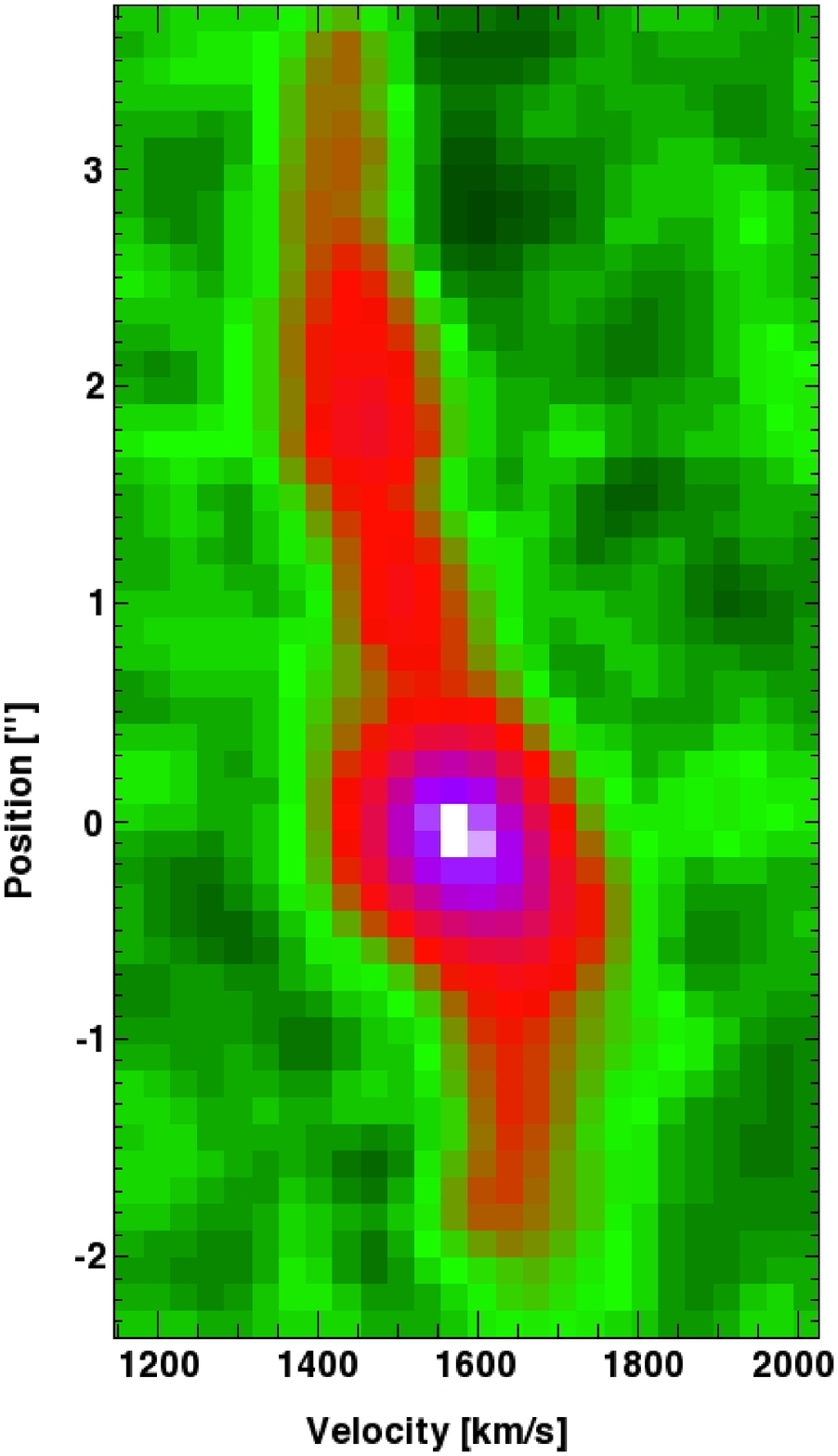}
\caption{H$_{2}$ 1$-$0 S(1) position - velocity diagram from north to
south, at the position of the compact \hh\ source.  The peak
corresponds to the compact \hh\ source (Sect. \ref{sec:cs}). Positions are
relative to this source.}\label{fig:pv}
\end{figure}

In the SINFONI data cube we identify several emission lines from
molecular (H$_{2}$~1$-$0~S(0)$\rightarrow$S(3), \hh~2$-$1~S(1)), and
ionized (\brg, \ion{He}{i}) gas. 
Fig.~\ref{fig:flx} shows the maps of emission-line surface
brightnesses, velocities, and FWHMs for H$_2$ 1$-$0 S(1) and
\brg. A position-velocity diagram is shown in Fig.~1. 
Mean velocity and line widths were measured with Gaussian
fits, carefully masking bad pixels. We also constructed a line-free
continuum map (lower right panel of Fig.~\ref{fig:pos}) by averaging
the flux across the full band in each spatial pixel after masking
emission lines, bad pixels, and wavelengths affected by strong
night-sky lines. 

\begin{figure*}[ht]
\centering
\includegraphics[width=18cm]{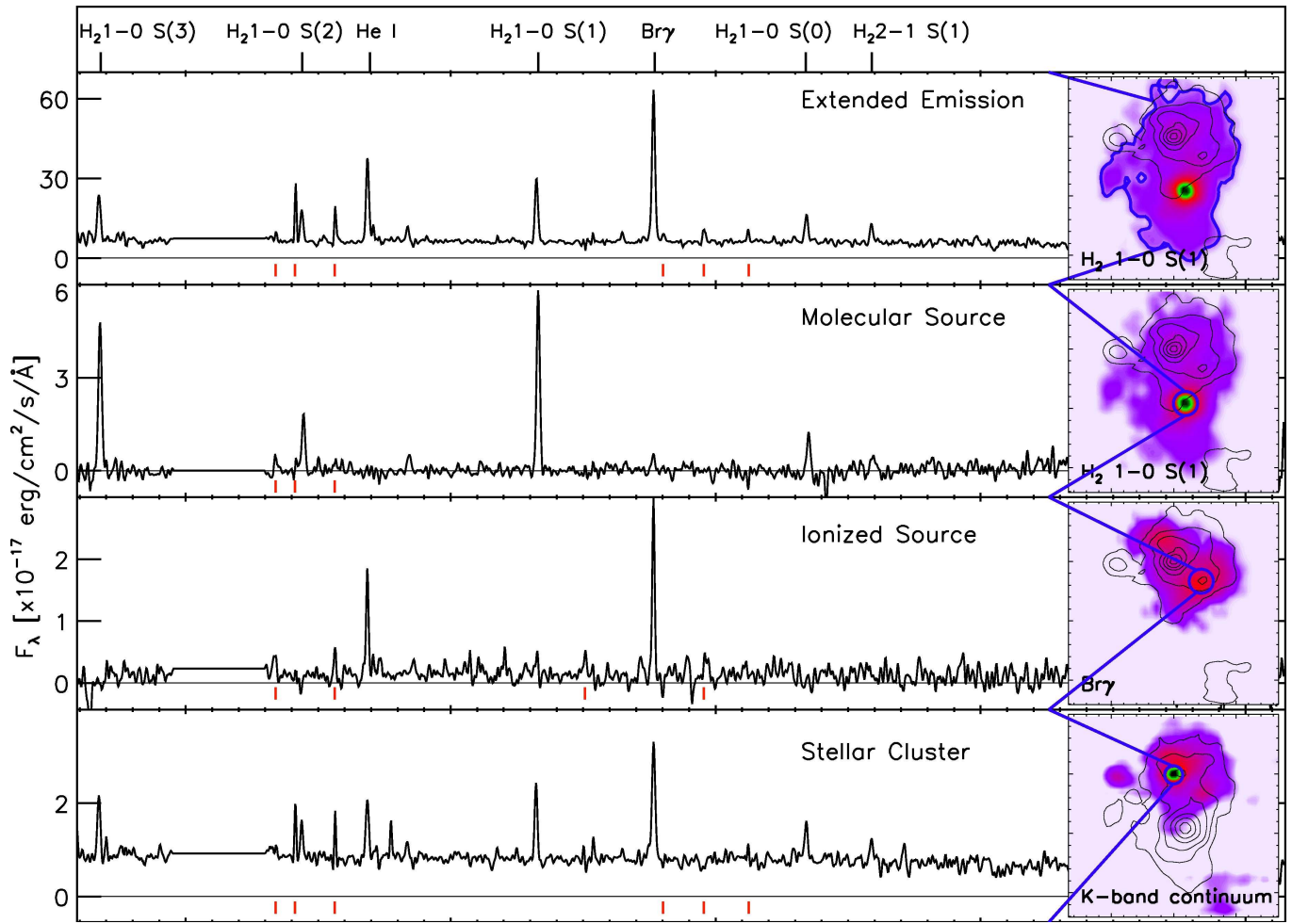}
\caption{Integrated SINFONI spectra for the four different components
in our pointing of the overlap region of the Antennae. All spectra are
smoothed by 3 bins in wavelength. Around $2.0~\mu$m the atmospheric
transmission is very low so we did not plot this part of the
spectra. We mark in red the wavelengths of the strong night sky lines
seen in the spectra (OH lines). In the right panel, blue contours mark
the aperture from which each spectrum was extracted.}\label{fig:spec}
\end{figure*}

We identify 4 components in these maps: a compact continuum source
(hereafter the super star cluster, SSC), two compact emission line
sources (hereafter the molecular, and the ionized compact source,
respectively) and diffuse extended emission.  The three compact
sources are labelled in Fig.~\ref{fig:spec}.  The SSC has previously
been identified with HST and Keck. The main parameters of the cluster
are listed in Table~\ref{tab:sc}. We estimated the stellar mass to $6
\tten{6} ~$\msun\ from the $K$-band and \brg\ fluxes corrected for
extinction (see Sect. \ref{sec:lum_ext}), using Starburst99
\citep{leitherer99}, assuming solar metallicity, an instantaneous
burst of age $4.5~$Myr, and a Salpeter initial mass function with
lower and upper mass cutoffs of 1 and 100~\msun, respectively.  Our
upper age limit of $6~$Myr is based on the absence of CO band-heads in
the K-band spectrum and is consistent with previous measurements
(Table~\ref{tab:sc}).  This cluster is one of the most massive
clusters in the overlap region, and one of the brightest in the
near-IR. The characteristics of the molecular and ionized compact
sources are presented in Table~\ref{tab:cs}.

The integrated spectra of each of the four components are presented in
Fig.~\ref{fig:spec}.  In Table~\ref{tab:mol}, we list line fluxes,
mean velocities and line widths for the extended emission, and the
molecular and ionized compact sources.  To construct the spectrum of
the extended emission, we integrated over the full spatial extent
where H$_2$ 1$-$0 S(1) emission is observed, and subtracted the
contribution from the compact sources. For the compact sources, we
integrated within circular apertures of $0\farcs6$, $0\farcs8$, and
$0\farcs7$ (for the stellar cluster, the molecular and ionized
sources, respectively) matched to their sizes. For the molecular and
ionized sources we subtract the nearby background using an outer
annulus.  For the SSC, we do not see a distinct emission-line
component that could be separated from the surrounding line emission.
Therefore, we did not subtract background line emission from the
spectrum of the SSC. Had we done so, we would have obtained a pure
continuum spectrum with no evidence of emission lines (not shown).

The spectra suggest that the astrophysical nature of these sources
must be very different.  For the extended line emission, we observe
emission lines from both molecular and ionized gas. The molecular
source shows only \hh\ lines and the ionized source only the \brg\ and
\ion{He}{i} emission lines.  Fig.~\ref{stages} presents a zoom onto
the \brg\ and H$_{2}$ 1$-$0 S(1) lines for the extended emission-line
region and the compact molecular source to illustrate the differences
in line intensities and kinematics. For the extended emission, the
velocities of the molecular and ionized gas are offset by $\sim100$ km
s$^{-1}$.  Throughout the rest of the paper we focus on the molecular
emission, i. e., the \hh\ extended emission and the compact \hh\
source. The analysis of the ionized gas will be presented in a future
publication.

\subsection{Extended molecular emission}\label{sec:ee}

\begin{figure}[ht]
\centering
\includegraphics[width=8.5cm]{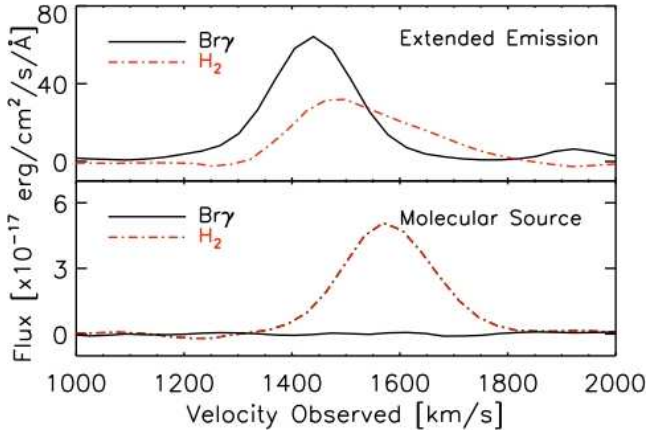}
\caption{Emission line profiles of molecular and ionized gas.  Solid
lines correspond to \brg\ and dot-dashed lines to H$_{2}$ 1$-$0
S(1). The kinematics of the extended molecular and ionized gas are
different with an offset of $\sim100$ km s$^{-1}$. The molecular
source shows broad \hh\ line emission and no \brg\
emission.}\label{stages}
\end{figure}

Fig.~\ref{fig:flx} shows extended H$_{2}$ line emission over most of
the field-of-view, with a morphology that is very different from that
of \brg\, which is more concentrated around the SSC.  The velocity
field shows a systematic gradient of $\sim200$~km~s$^{-1}$ from
north-east to south-west over a projected distance of $600$~pc, giving
an average velocity gradient of
0.35~km~s$^{-1}$~pc$^{-1}$. Figure~\ref{fig:pv} shows the position -
velocity diagram for the H$_{2}$ 1$-$0 S(1) line, along the direction
of the velocity gradient (P.A. 15$^{ \circ} $). The diagram is
centered on the compact \hh\ source and reveals the velocity gradient
and the high turbulence in the field.  The H$_{2}$ line widths of
${\rm FWHM =150-200~km~s^{-1}}$ are comparable to the total gradient
over the SINFONI field-of-view. The H$_{2}$ line width is larger than
those measured in \brg\ for the ionized gas (Fig.~\ref{stages}).

The \hh\ rotational lines observed with Spitzer \citep{brandl09} are
not spectrally resolved, the highest spectral resolving power of these
data is ${\rm R=600}$, corresponding to ${\rm
FWHM=500\,km\,s^{-1}}$. The CO(1$-$0) velocity ($V_{\rm LSR}=1470$~km~s$^{-1}$)  agrees with the mean,
luminosity weighted, velocity of the \hh\ extended emission
(Table~\ref{tab:mol}). CO shows a velocity gradient as large as that
of \hh\ but over larger distances. CO observations with a higher
angular resolution are needed to resolve this gradient at the same
scale as H$_2$. The warm H$_{2}$ gas traced by the near-IR emission
lines has line widths more than twice as large as those measured in CO
at the same position \citep[73 km~s$^{-1}$ in][]{wilson00}.

\begin{table*}[ht]
\caption{Emission-line characteristics.}\label{tab:mol}
\begin{center}
\begin{tabular}{p{1.55cm} c c c l c c l c c c}
\hline \hline
\noalign{\smallskip}
\multirow{3}{1.55cm}{\hspace{5mm}Line}         &   \multirow{2}{1.5cm}{\hspace{5mm}Rest Wavelength}      &\multicolumn{3}{c}{Extended Emission} & \multicolumn{3}{c}{Molecular Compact Source} & \multicolumn{3}{c}{Ionized Compact Source}\\[0.5ex]
\cline{3-11} \\[-1.5ex]
       &     &V$_\mathrm{LSR}$ & Flux$\tten{15}$    & FHWM$^{a}$ & V$_\mathrm{LSR}$ & Flux$\tten{16}$     & FWHM  & V$_\mathrm{LSR}$ & Flux$\tten{17}$ & FWHM      \\

\hline 
\noalign{\smallskip}
H$_{2}$ 1$-$0 S(3)  &   \hspace{2mm}1.95756 &  1490$\pm$17 &  6.1$\pm$0.4 & 225$\pm$14 & 1515$\pm$16 & 6.5$\pm$0.2 & 167$\pm$7    &  &   & \\
H$_{2}$ 1$-$0 S(2)  &   \hspace{2mm}2.03376 &  1461$\pm$18 &  3.3$\pm$0.3 & 234$\pm$18 & 1538$\pm$17 & 2.3$\pm$0.1 & 140$\pm$11  &   &  & \\
H$_{2}$ 1$-$0 S(1)  &   \hspace{2mm}2.12183 &  1493$\pm$17 &  6.9$\pm$0.3 & 215$\pm$9   & 1547$\pm$16 & 7.6$\pm$0.1 & 146$\pm$2    & 1481$\pm$18 &    3.1$\pm$0.6 & $<$136\\
H$_{2}$ 1$-$0 S(0)  &   \hspace{2mm}2.22329 &  1465$\pm$18 &  2.8$\pm$0.2 & 188$\pm$16 & 1554$\pm$18 & 1.6$\pm$0.2 & 140$\pm$22  &  &   & \\
H$_{2}$ 2$-$1 S(1)  &   \hspace{2mm}2.24772 &  1450$\pm$18 &  1.7$\pm$0.2 & 146$\pm$20 & 1530$\pm$26 & 0.9$\pm$0.2 & 243$\pm$44  &   & $<$0.36 & \\
\hspace{4mm}\brg\ & \hspace{2mm}2.16612 &  1408$\pm$16 &  9.5$\pm$0.2 & 133$\pm$4   & 1443 &   $<$0.06      &                       & 1404$\pm$16 &  25.8$\pm$1.1 & 20$\pm$29 \\
\hspace{4mm}He I  & \hspace{2mm}2.05869 &  1410$\pm$16 &  5.2$\pm$0.2 & 131$\pm$7  &                  &                      &                       & 1400$\pm$16 &  16.3$\pm$1.0 & 35$\pm$27 \\
Continuum$^{b}$    &                                      &           &         0.06        &        &   &       &         &   &  0.1 &   \\[0.3ex]
\hline
\end{tabular}

\end{center}
Wavelengths are given in $\mu$m, fluxes in \uflux, and velocities and
FWHM in km~s$^{-1}$. Error bars include the uncertainties in the
wavelength calibration.\\ $^{a}$ intrinsic FWHM after deconvolution
with the instrumental resolution of $\sim$136~km~s$^{-1}$,
$^{b}$ line-free $K$-band continuum flux in units of \uflux
$\AA^{-1}$.
\end{table*}

\subsection{Compact molecular source} \label{sec:cs}

The \hh\ flux map in Fig.~\ref{fig:flx} reveals a compact source
without a counterpart in the ionized gas and continuum emission.  The
spectrum of this source in Fig.~\ref{fig:spec} and \ref{stages} is
unlike that of the other emission components. To our knowledge, this
is the first time that an extragalactic source with a $K$-band
spectrum showing only H$_{2}$ lines is discovered. Similar spectra are
seen in the Milky Way towards low-mass stars embedded in molecular
clouds, which do not produce enough ionizing photons to power bright
Hydrogen recombination lines \citep{giannini06}. In these sources the
H$_{2}$ emission traces shock heating of H$_{2}$ associated with the
interaction of the stellar jet with the molecular cloud.

Table~\ref{tab:mol} gives the H$_{2}$ line fluxes and an upper limit
on the \brg\ flux. The \hh~1$-$0~S(1) flux of the molecular compact
source accounts for $10$\% of the total emission from SGMC~2 in this
line.  To estimate the size of this source, we fit the
azimuthally-averaged emission-line surface brightness profile along
right ascension and declination with Gaussian curves. The FWHM size,
geometrically averaged over the two axes, is $0\farcs8$. Correcting
for the seeing we obtain an intrinsic source size of $0\farcs5$
corresponding to $50$~pc. 

\begin{figure}\centering
\includegraphics[width=9cm]{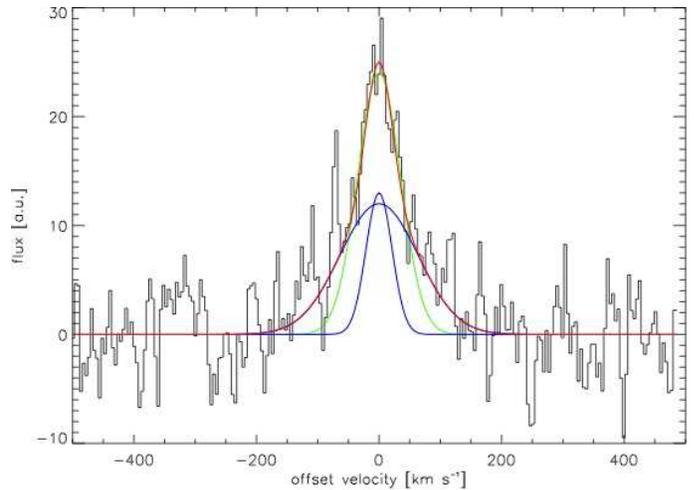}
\caption{Integrated spectra of the compact \hh\ source observed with
CRIRES. Fitting the data with a single Gaussian (green line) shows
significant residuals. An adequate fit (red line) requires 2 Gaussian
curves (blue lines), corresponding to a narrow and a broad component,
respectively.}
\label{fig:cri}
\end{figure}

We obtained high spectral resolution measurements of the line profile
of the compact source with CRIRES (Fig.~\ref{fig:cri}). Fitting this
profile requires two Gaussian components with different line widths at
a common velocity, with FWHM=50 km s$^{-1}$ and FWHM=160 km s$^{-1}$,
respectively. We illustrate this by showing fits with one and two
Gaussians in Fig.~\ref{fig:cri}. In the two-Gaussian fit, the narrow
component contributes 25\% to the total \hh\ line flux. We verified
that we recover the line profile measured with SINFONI after
convolution with the line spread function of SINFONI.

The compactness of the H$_{2}$ source suggests that it is
gravitationally bound, and that we can estimate the gas mass using the
virial theorem.  For a homogeneous spherical cloud the balance between
gravitational and kinetic energy gives the virial mass as
$M_\mathrm{vir}=5~R~\sigma^{2}/G$, where $R$ is the radius, $\sigma$
the velocity dispersion of the gas, and $G$ the gravitational
constant. In this formula, we estimate the diameter using the FWHM
size of the near-IR \hh\ emission.  We obtain a virial mass of
$1.3\tten{7}\times (\sigma/21\,{\rm km~s^{-1}})^2$~\msun\ using
the line width of the narrow component detected with CRIRES for $\sigma$. The CO
data do not have the required angular resolution to separate the
compact source from the extended emission. High angular resolution
observations with ALMA are required to estimate the virial mass from
CO.  The mass of the compact \hh\ source is a factor $\sim30$
smaller than the mass derived from CO observations for the SGMC 2
molecular complex \citep{wilson00}. From the virial mass and size, we
derive the column density of the molecular source and we obtain
$N_\mathrm{H} = 6\tten{23}$~H~cm$^{-2}$. This corresponds to a mean density of 6000 H
cm$^{-3}$. 

Radio continuum maps at 6~cm show a slight excess (50 $\mu$Jy) at the
position of our molecular source \citep[see Figures 5 and 9
in][]{neff00}.  Assuming that this radio flux is entirely thermal
emission from ionized gas, it corresponds to a \brg\ flux of $\sim 5
\times 10^{-16}$~\uflux.  This value is a factor 5 larger than the
upper limit on the observed \brg\ flux in Table~\ref{tab:mol} after
scaling by our estimate of the extinction correction (a factor 20, see
Table~\ref{tab:cs} and Sect. \ref{sec:lum_ext}).  This slight discrepancy
could indicate that the newly formed stars are concentrated at the
center of the cloud.  In this case their emission could be more
attenuated than the \hh\ emission.  From the radio flux, we compute an
ionizing photon rate of $\rm 2\times 10^{51} \, s^{-1}$, which, for a
young (1-2~Myr) burst corresponds to a stellar mass of $\sim 4 \times
10^4~$\msun\, based on Starburst99 models.  This stellar mass is 0.3\%
of the virial mass of the compact \hh\ source.


\begin{table}[ht]
\caption{Observed characteristics of the compact sources.}\label{tab:cs}
\begin{center}
\begin{tabular}{l c c}
\hline \hline
\noalign{\smallskip}
Parameters & H$_{2}$ source & \ion{H}{ii} source \\
\hline
\noalign{\smallskip}
Size                                                 & 0\farcs5                    & 0\farcs4\\ 
$e_\mathrm{c}$                              & $\sim$20                   & $-$ \\
 $n_\mathrm{e}$ [cm$^{-3}]$          & $-$                            & 58 $^{a}$ \\
Mass  [\msun]                                  & 1.3$\tten{7}$ $^{b}$ & 4.8$\tten{4}$ $^{c}$ \\
 $N_\mathrm{ion}$ [phot. s$^{-1}$] &  $<$2.6$\tten{49}$ $^{d}$     & 8.5$\tten{50}$ $^{e}$\ \\
$L_\mathrm{bol}$  [\lsun]                           & 2.6$\tten{6}$  $^{f}$   & 5$\tten{7}$ $^{g}$\\ 
\hline
\end{tabular}

\end{center}

$^{a}$ assuming ionization balance, and not correcting for extinction, scales as $e_\mathrm{c}^{0.5}$. 

$^{b}$ virial mass estimated from the intrinsic S(1) size and the CO line width. 

$^{c}$ mass of the \ion{H}{ii} gas uncorrected for extinction, scales as $e_\mathrm{c}^{0.5}$. 

$^{d}$ from the upper limit of the Br$\gamma$ emission, uncorrected for extinction.

$^{e}$ derived from the Br$\gamma$ luminosity, scales as $e_\mathrm{c}$.

$^{f}$ H$_{2}$ bolometric luminosity corrected for extinction (Sect. ~3.3).

$^{g}$ from the age and $N_\mathrm{ion}$. Uncorrected for extinction, scales as $e_\mathrm{c}$.

\end{table}

\section{The nature of the compact and diffuse  H$_{2}$ emission}\label{sec:nat}

In this section we argue that, for both the extended emission and the
compact molecular source, the H$_2$ line emission is shock powered. We
also estimate the bolometric H$_{2}$ luminosity for both emission
components.

\subsection{Spectral diagnostics: PDR or shocked gas? }\label{sec:shock}

In this section, we discuss the excitation of the H$_2$ gas and the
nature of the H$_{2}$ emission for the extended emission and the
compact source. We focus on the near-IR lines within the $K$-band
using spectral diagnostics that do not depend on extinction.

\begin{table}[ht]
\caption{\hh\ line emission: Spectral diagnostics}\label{tab:sd}
\begin{center}
\begin{tabular}{l c c}
\hline \hline
\noalign{\smallskip}

Component & \hh\  2$-$1/1$-$0~S(1) & \hh\ 1$-$0~S(1)/Br$\gamma$    \\
\hline
\noalign{\smallskip}
Extended emission                &     &                 \\ 
\hspace*{0.5cm}Full area                   &    0.2      & 0.72 \\ 
\hspace*{0.5cm}Southern part                     &             & 3 \\
 \hh\ compact source            & 0.1   & $>$100\\  
\hline
\end{tabular}

\end{center}

\end{table}

The H$_{2}$ excitation diagrams for the extended emission and the
compact H$_{2}$ source are presented in Fig.~\ref{fig:exc}. We obtain
two estimates of the gas temperature, the first by fitting the H$_{2}$
population in the rotational states of the v=1 vibrational state, and
the second from the ratio between the J=1 level of the v=1 and v=2
states. For the compact H$_{2}$ source we find $1700$~K and $2300$~K,
respectively, and for the extended emission $900$~K and $2700$~K. The
molecular gas has a range of temperatures. Most of the molecular gas
is too cold to emit in the near-IR. The mass of warm gas seen in the
near-IR is several orders of magnitude smaller than that observed in the mid-IR rotational lines by
\citet{brandl09}, and also much smaller
than the virial mass given in Sect. \ref{sec:cs}. However, even though the
mass is small, the energies are very large as we will see in
Sect. \ref{sec:lum}.

The ratio $R_{2-1/1-0}\equiv$~H$_2$~2$-$1~S(1)/H$_2$~1$-$0~S(1) is
often used to discriminate between UV heating of H$_{2}$ in
photodissociation regions (PDRs) and shock heating.  The observed
values of the extended emission and compact molecular source are $0.2$
and $0.1$, respectively. We compare these values with the Meudon PDR
code described in \citet{lepetit06}. Their Fig. 27 shows that PDRs
typically have $R_{2-1/1-0}=0.5-0.6$. Values of $R_{2-1/1-0}$ in the
range 0.1 to 0.2 are only reached for very high densities and strong
radiation fields ($n\gtrsim10^{5}$~cm$^{-3}$,
$\chi\gtrsim10^{5}$). Such conditions do exist in massive star-forming
regions, but only close to massive stars. This solution does not apply
here because the compact \hh\ source is far away from the SSC, and has
no strong ionized line emission indicating the absence of massive
young stars.  For the compact source PDR emission is also ruled out by
the ratio H$_2$~1$-$0~S(0)/H$_2$~1$-$0~S(1) \citep[see Fig.~28
of][]{lepetit06}.  The extended \hh\ emission-line region extends over
the whole field-of-view and does not peak at the position of the
SSC. We therefore discard UV heating and favor shocks as main heating
mechanism for the extended emission as well. Using the models of
\citet{kristensenphd}, we find that J- and C-shocks match the observed
values over a wide range of densities ($10^{4} - 10^{6}$~cm$^{-3}$)
and shock velocities ($15-50$ km~s$^{-1}$).

\begin{figure}[ht]
\centering
\includegraphics[width=5.5cm]{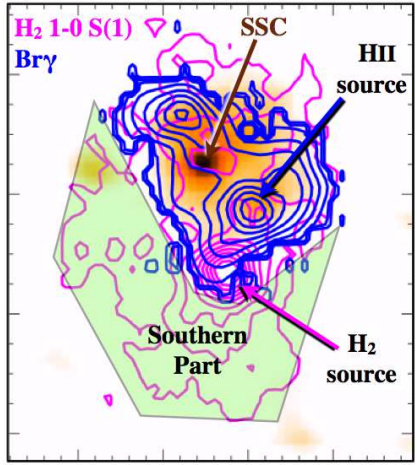}
\caption{The color image shows the $K$-band continuum, which peaks at the SSC.
Pink and blue contours show the \hh\ 1$-$0 S(1) and \brg\ emission, respectively. 
We also mark the compact ionized and molecular source (\ion{H}{ii} source and \hh\
source, respectively). The green area corresponds to the gas which is likely not
influenced by the SSC, we call it 'Southern part'. }\label{fig:cont}
\end{figure}

\begin{figure}[ht]
\centering
\includegraphics[width=8.7cm]{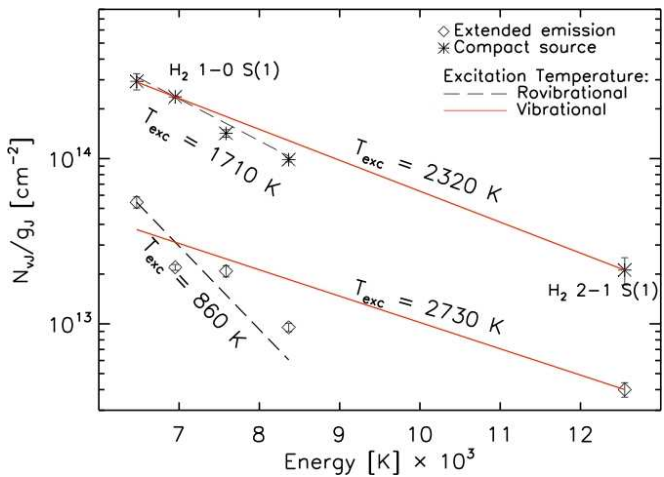}
\caption{H$_{2}$ excitation diagram of the extended emission-line
region and the compact source (Sect. ~3.1). The dashed and solid red lines show
the gas excitation temperature derived from fitting the H$_{2}$
population in the rotational states of the v=1 vibrational state, and
from the ratio between the J=1 level of the v=1 and v=2 states,
respectively.}\label{fig:exc}
\end{figure}

The ratio $R_\mathrm{S(1)/Br\gamma}\equiv$~H$_2$~1$-$0~S(1)/Br$\gamma$
is an additional, more empirical diagnostics to identify the origin of
the H$_{2}$ excitation.  These diagnostics provide additional
evidence of shock excitation of H$_{2}$ in our data.  We compare our
results for the Antennae overlap region with data presented by
\citet{puxley90}. Their data include 44 star-forming regions in 30
galaxies, where, typically, the H$_2$~1$-$0~S(1) line is weaker than
the Br$\gamma$ line (only 4 regions have a ratio higher than 1). Their
flux ratios range between $0.1$ and $1.5$ with a mean of $0.5$ and a
dispersion of $0.3$.  A few sources such as the merger NGC~6240 show a
much higher $R_\mathrm{S(1)/Br\gamma}$ value.  \citet{puxley90}
present models for a number of different scenarios and conclude that
the line emission is most likely from \ion{H}{ii}/PDR gas in massive
star forming regions, for most of their sources.  With our data, we
obtain $R_\mathrm{S(1)/Br\gamma}=0.72$ and $>100$ for the extended
emission and the compact source, respectively. The compact source
shows a very high ratio, which is far from the typical range in
star-forming galaxies. The ratio of the extended emission is within
the range of observed and modeled values by \citet{puxley90}. However,
this is not true for all of the extended gas. In the region South of
the bright \brg\ line emission (Fig.~\ref{fig:cont}), we measure a
ratio of 3. This value suggests that at least parts of the extended
H$_2$ emission is excited by shocks.

\subsection{H$_{2}$ bolometric luminosity}\label{sec:lum}

Before discussing the H$_2$ emission as a coolant of the molecular gas
in Sect. \ref{sec:large}, we need to estimate the H$_2$ bolometric
luminosity of the extended emission and the compact source. This will
be used in Sect. \ref{sec:dis} and Sect. \ref{sec:ssc_form} to discuss what
may power the \hh\ emission. To achieve this we combine the Spitzer
and SINFONI H$_2$ line measurements.  This is not straightforward to
do because the near-IR H$_2$ is attenuated by dust extinction and the
Spitzer angular resolution is too low to separate the contributions of
the extended and compact source to the mid-IR rotational line
emission.  First, we estimate the near-IR extinction for both emission
components (Sect. \ref{sec:lum_ext}). Second, we estimate the bolometric
H$_2$ luminosities, under the assumption that the fraction of the
H$_{2}$ luminosity in the H$_2$ 1$-$0 S(1) line is the same for the
extended emission and the compact source (Sect. \ref{sec:lum_bol}).

\subsubsection{Extinction correction}\label{sec:lum_ext}
To quantify the extinction in the extended gas, we compare the number
of ionizing photons $N_\mathrm{ion}$ obtained from the observed
Br$\gamma$ emission line with that computed from the fluxes in the
mid-IR [\ion{Ne}{ii}]$\lambda$12.8$\mu$m and
[\ion{Ne}{iii}]$\lambda$15.6$\mu$m fluxes \citep{brandl09}.  The
extinction corresponds to the ratio between the two estimated values
of $N_\mathrm{ion}$. This ratio provides the effective extinction
factor averaged over the entire field-of-view at the wavelength of
\brg\ because the dust extinction of the mid-IR neon lines is much
smaller than in the near-IR. The observed $N_\mathrm{ion}$ in the
field-of-view was measured by integrating the \brg\ flux over the
entire SINFONI cube. Assuming case B recombination, an electron
density $n_\mathrm{e}=10^{2}$~cm$^{-3}$ and an electron temperature of
$T_\mathrm{e}=10^{4}$~K \citep[see Table~8 of ][]{hummer87},
$N_\mathrm{ion}=4.4\tten{52}$ phot. s$^{-1}$. To compute the
$N_\mathrm{ion}$ from the Ne fluxes, we use the equations in
\citet{ho07} and the [\ion{Ne}{ii}]$\lambda$12.8$\mu$m and
[\ion{Ne}{iii}]$\lambda$15.6$\mu$m fluxes. We further assume that $n_\mathrm{e}$ is smaller
than the critical density,$\sim10^{5}$ cm$^{-3}$, and adopt the
galactic Neon abundance. We obtain $N_\mathrm{ion}=6.6\tten{52}$
phot. s$^{-1}$. Comparing with the result for \brg\, this suggests an
extinction factor of 1.5.

The compact source has a very high column
density, $N_\mathrm{H}=6\tten{23}$ H cm$^{-2}$ (Sect. \ref{sec:cs}), and
therefore, the extinction must be much higher. Using the Milky Way
extinction curve for $R_{V}=3.1$ in \citet{weingartner01} this column
density corresponds to $\tau_\mathrm{ext}=30$ in the $K$-band. We 
roughly estimate the extinction correction from
$\tau_\mathrm{ext}(K)$ with a simple toy model. We compute the
attenuation due to the dust absorption for every point of a spherical
and homogeneous cloud of constant density and line emission per unit
volume. The flux that reaches the surface of the cloud depends on the
path length $l$, and is attenuated by a factor
e$^{-\tau_{\lambda}(l)}$. The optical depth
$\tau_{\lambda}(l)=\kappa_{\lambda}\,\rho\times l$ depends on the
absorption coefficient at a given wavelength $\kappa_{\lambda}$ and
the mass density $\rho$. The $\kappa_{\lambda}$ value is taken from
the $R_{V}=3.1$ curve in \citet{weingartner01}. We sum the attenuated
emission over the cloud and compare the result with the total emission
for zero opacity. We find a correction factor $e_\mathrm{c}=0.7\times
\tau_\mathrm{ext}$. We also run the model with a density and emission
profile $\rho \propto r^{-2}$ for which we find
$e_\mathrm{c}=0.6\times \tau_\mathrm{ext}$. In the following we only
apply this extinction correction ($e_\mathrm{c}=20$) to the narrow
component of the \hh\ line emission. In our interpretation of the
data, the broad component is likely to come from the surface of the
cloud, for which the extinction correction should be about a factor of
2, because we do not see the emission behind the cloud.

\subsubsection{Bolometric correction }\label{sec:lum_bol}
The angular resolution of Spitzer/IRS is only 5\arcsec, too low to
measure the total H$_{2}$ luminosity, H$_2^\mathrm{bol}$, including
the pure-rotational mid-IR lines, of the compact source directly. To
circumvent this difficulty, we need to make an assumption.  We assume
that the H$_2^\mathrm{bol}$ /H$_2$~1$-$0~S(1) ratio, after extinction
correction, is the same for the compact source and for the extended
emission.  The detailed shock modelling which would be needed to
support or refine this assumption is beyond the scope of this
observational paper.  The total H$_{2}$ luminosity (extended and
compact source) is taken from the observations by \citet{brandl09},
assuming that most of the H$_{2}$ emission is from the first four
rotational lines.  Within these simplifying assumptions, we find
bolometric H$_{2}$ luminosities of $L_\mathrm{H_{2}} = 5.4\tten{6}$
and $2.6\tten{6}$ \lsun\ for the extended and compact emission,
respectively.

The extinction correction introduces significant uncertainties to
these luminosities, in particular for the compact source. The total
\hh\ luminosity (extended emission + compact source) does not depend
on extinction because most of the \hh\ emission is from the mid-IR
v=0 rotational lines measured by Spitzer for which extinction is
negligible. It is the ratio between both luminosities which depends on
the extinction correction. We set a lower limit of $8\tten{5}$~\lsun\
on the luminosity of the compact source by assuming that the
extinction is the same for both components. The \hh~1$-$0~S(1)
emission represents $3$\% of the bolometric \hh\ emission.


\section{From large scale gas compression to star formation}\label{sec:large}

The results presented in the previous sections raise three questions.
How were the extended molecular gas component and the compact \hh\
source formed? Why is this gas so \hh\ luminous?  What is the nature
of the compact \hh\ source?  In this section, we argue that the
extended emission traces highly turbulent molecular gas formed where
the galaxy interaction is driving a large scale convergent flow. We
relate the H$_2$ emission to the dissipation of the gas turbulent
kinetic energy and the formation of the SGMC~2 complex and the compact
\hh\ source by gas accretion (Sect. ~5.2). In Sect. \ref{sec:ssc_form} we argue that
the compact \hh\ source is a massive core on its way to forming a
super star cluster.

\subsection{Gas compression and gravitational fragmentation}\label{sec:gf}

The Antennae merger is close to pericenter passage when tidal forces
are compressive \citep{renaud08,renaud09}.  Thus, we consider that the
velocity gradient observed with the H$_2$ line emission across the
SINFONI field is tracing a convergent flow driven by the galaxy
interaction. The associated gas compression has created the conditions
for the gas to cool and to become molecular, like in numerical
simulations of convergent flows \citep{hennebelle08}. As observed in
the galaxy collision in Stefan's Quintet, the mechanical energy of the
interaction is not fully thermalized in large scale shocks
\citep{guillard09}. Due to the inhomogeneous, multi-phase, nature of
the interstellar medium much of the gas kinetic energy decays from
large to small scales before it is dissipated.  Thus, we consider that
it is the galaxy interaction that drives the molecular gas turbulence.
This interpretation is supported by the fact that the large velocity
gradient has the same magnitude as the turbulent line width, and by
numerical simulations, which illustrate how dynamical and thermal
instabilities lead to the formation of highly structured and turbulent
molecular clouds where gas flows collide
\citep{heitsch05,vazquez07,hennebelle08}.

The gas surface density is high enough for the turbulent gas to
gravitationally contract and fragment within the time scale over which
the tidal forces are compressive \citep[$\sim$10 Myr,][]{renaud08}.
We compute the Jeans length ($R_\mathrm{Jeans}=5\sigma^{2}/(2\pi G
\sigma)$) and mass ($M_\mathrm{Jeans}= \pi \,
R_\mathrm{Jeans}^{2}\sigma$) using the CO(1$-$0) observations of
\citet{wilson00} to estimate the gas velocity dispersion ($\sigma$)
and surface density ($\sigma$). The value of $\sigma$ is
30~km~s$^{-1}$.  The surface density $\sigma$ is obtained taking the
ratio between the total virial mass and the total area of the overlap
region $\sigma\sim500$~\msun~pc$^{-2}$.  We find
$R_\mathrm{Jeans}\simeq 350$~pc and $M_\mathrm{Jeans} =
2\tten{8}$~\msun.  To compute the free-fall time scale \tff\
$=\sqrt{3\pi/(32\rho G)}$ we determine the mean density $\rho$ from
$R_\mathrm{Jeans}$ and $M_\mathrm{Jeans}$. We find \tff\ $=$ 8~Myr.

The Jeans length and mass are comparable to the size of the super
giant molecular complex SGMC~2.  It is thus likely that the complex
has formed by gravitational contraction out of the gas compressed by
the galaxy interaction.  The gravitational contraction could
contribute to the H$_2$ velocity gradient. An estimate of the motions
driven by the self gravity of the complex is obtained taking the ratio
between the free-fall velocity
$v_\mathrm{ff}=2~R_\mathrm{Jeans}/\tau_\mathrm{ff} \simeq 90$
km~s$^{-1}$ and the Jeans radius.  We find a value comparable to the
velocity gradient measured with the H$_2$ line emission (200 km
s$^{-1}$ over 600 pc). This calculation provides an upper limit to the
velocity gradient from the cloud contraction since it is unlikely that
the complex is free-falling. The cloud contraction must occur at a
slower rate than free-fall because it takes time to dissipate
turbulent energy \citep{elmegreen07,huff07}.  

Thus, we propose that the SGMC~2 complex is formed by convergent gas
flows driven by the galaxy interaction and the gas self-gravity.
\citet{wilson00} find that the gas mass inferred from the CO line
luminosity matches the virial mass estimated from the the CO line
width and conclude that the SGMC~2 complex is gravitationally
bound. In the remainder of this paper we use 'CO complex' to refer to
gravitationally bound gas in the SGMC~2 complex.
The velocity gradient and line width of the H$_2$ emission are both
larger than the CO line width, suggesting that not all gas in the
SGMC~2 complex is gravitationally bound. Most of the H$_2$ emission
arises from gas that is unbound.

In the next two sub-sections we will interpret our H$_2$ data within
this scenario. We associate the H$_2$ emission with the formation of
the CO complex through gas accretion. We argue that the H$_2$ emission
traces the dissipation of kinetic energy, which is required for gas
that is driven by the galaxy interaction to become gravitationally
bound.

\subsection{The energetics and formation of the SGMC~2 complex}\label{sec:dis}

From observations of the Milky Way \citep{falgarone05}, and, more
recently, of extragalactic sources
\citep{guillard09,nesvadba10,ogle10}, we know that H$_2$ line emission
is a major coolant of the ISM associated with the dissipation of
interstellar turbulence. The H$_2$ line emission may arise from
shocks, as quantified by models such as \citet{flower10}, and from
friction betweens ions and neutral species in vortices
\citep{godard10}.  Models show that other emission lines contribute to
the gas cooling but that they do not change the order of magnitude of
the cooling rate \citep{flower10,godard10}. Based on this earlier
work, we consider the extended \hh\ emission from SGMC~2 as a
quantitative tracer of the dissipation rate of the kinetic energy of
the gas.

Since the H$_2$ line widths and the velocity gradient across
SGMC~2 is twice the CO line width, the H$_2$ emission does not come
solely from the dissipation of the turbulent kinetic energy of the CO
complex.  A significant fraction of the emission must come from the
dissipation of bulk and turbulent kinetic energy of gas driven
by the galaxy interaction. Such a loss of kinetic energy is a required
step for this gas to become gravitationally bound.

In the following, we quantify this interpretation which connects the
energetics of the \hh\ gas to the formation of the CO complex by gas accretion.
We consider that the CO complex evolves in a quasi-static way assuming
that virial equilibrium applies at all times.  The virial equation
relates the gravitational and turbulent kinetic energy,
$E_\mathrm{grav}$ and $E_\mathrm{turb}$, and includes the external
pressure $P_\mathrm{ext} $:

\begin{equation}\label{eqV}
E_{\rm grav} + 2\, E_{\rm turb} =  3\, P_{\rm ext} \, V 
\end{equation}

\noindent
where $V$ is the volume. The turbulent energy is $E_{\rm turb} = 3/2 \,
M_{\rm CO} \, \sigma_{\rm CO}^2$, where $M_{\rm CO}$ is the mass of
the CO complex and $\rm \sigma_{CO}$ is the velocity dispersion along
the line of sight derived from the integrated CO spectrum of SGMC~2.
For a fixed $P_{\rm ext}$, the exchange of energy associated with gas
accretion and radiation is associated with the derivative of the
enthalpy $H$ of the complex \citep{huff07}:

\begin{equation}
\dot{H} = \dot{E}_{\rm grav} + \dot{E}_{\rm turb} + P_{\rm ext} \, \dot{V} =  -\dot{E}_{\rm turb} + 4 \, P_{\rm ext} \, \dot{V} = \dot{E}_{\rm in} -  L_{\rm H_2} / f_{\rm H_2}   
\label{eqH}
\end{equation}

\noindent
where $\dot{E}_\mathrm{in}$ is the energy input from gas accretion,
$L_\mathrm{H_2}$ the H$_2$ luminosity, and $f_\mathrm{H_2}$ the
fraction of the gas bolometric luminosity that is radiated in the
H$_2$ lines used to compute $L_\mathrm{H_2}$.  This fraction is
smaller than 1 because some cooling occurs in lines which have not
been measured.  Combining H$_2$ and far-IR observations,
\citet{maret09} find that $f_{\rm H_2}$ is in the range $\sim 0.25\, -
\, 0.5$ for shocks associated with gas outflow from low mass stars.
We assume that the kinetic energy of the accreted gas is the main
source of energy that balances radiative losses \citep{klessen10}.  In
doing this we neglect the energy that comes from stellar feedback. The
fact that the H$_2$ emission shows no enhancement around the SSC is an
indication that stellar feedback does not have a significant
contribution to the H$_2$ emission (Sect. \ref{sec:stef}).

We follow \citet{klessen10} in introducing the efficiency factor
$\epsilon $ which represents the fraction of the accretion energy that
drives turbulence. The turbulent energy dissipates over a time scale
$t_\mathrm{dis}({\rm CO}) \simeq R/\sigma_{\rm CO}$ where $R \simeq
300 \, {\rm pc}$ is the radius of the SGMC~2 complex
\citep{mckee07}. The balance equation between energy input
and dissipation is:

\begin{equation}\label{eqD}
\epsilon \times \dot{E}_{\rm in} =  3/2 \, M_{\rm CO} \times \sigma_{\rm CO}^3/R
\end{equation}

We compute the mass accretion rate necessary to drive the gas
turbulence by equating the energy input and the radiative losses. This
corresponds to a solution of equation (\ref{eqH}) where the derivative
of the gas kinetic energy $\dot{E}_{\rm turb}$ and the term associated
with $\dot{V}$ are both small with respect to the right-hand 
terms.  Thus, we write:

\begin{equation}\label{eqA}
\dot{E}_{\rm in} =  3/2 \, \dot{M}_{\rm acc} \, \sigma_{\rm H_2}^2 \simeq  L_{\rm H_2} / f_{\rm H_2}  
\end{equation}

\noindent
where $\sigma_{\rm H_2}$ is the velocity dispersion along the line of
sight derived from the H$_2$ data.  To quantify $\dot{M}_{\rm acc}$
and $\epsilon$, we consider only the turbulent component of the
velocity field ($\rm FWHM_{H_2} \sim 150 \, km \, s^{-1}$), assuming
that accretion occurs from turbulent gas with a mean velocity equal to
that of the CO complex. For $\rm \sigma_{H_2} = 65 \, km \, s^{-1}$
and $\rm \sigma_{CO} = 30 \, km \, s^{-1}$, we find $\dot{M}_{\rm acc}
= 5.3 / f_{\rm H_2}\,$\msun$\, \mathrm{yr^{-1}} $ and $\epsilon = 1.8
\times f_{\rm H_2}$.  For our interpretation to hold we must have
$\epsilon < 1$ , and thus $f_{\rm H_2} < 0.5$.  The fact that the
H$_2$ line width is larger than that of CO shows that this holds. In
the following, for numerical calculations, we consider that $f_{\rm
H_2} \sim 0.25$.

We scale $\dot{M}_{\rm acc}$ by the turbulence dissipation time scale
to estimate the mass of gas from which accretion is occurring: $M_{\rm
acc} = \dot{M}_{\rm acc} \times t_{\rm dis}({\rm H_2})$, where $t_{\rm
dis}(\mathrm{H_2}) \simeq R/\sigma_{\rm H_2}$ is the turbulent
dissipation time scale. The large scale flow can feed this mass
reservoir, compensating for what is accreted by the CO complex,
because the dissipation time scale is comparable to the dynamical time
scale associated with the velocity gradient across SGMC~2 ($3\times
10^6 \, $yr). We find $M_{\rm acc} = 2.4 \times 10^7 / f_{\rm
H_2}$~\msun. For $f_{\rm H_2} \sim 0.25$, $M_{\rm acc}$ is 25\% of
$M_{\rm CO}$.  Thus, the mean density in the CO complex may only be a
few times larger than that of the accreted gas. This agrees with our
estimate of the efficiency factor $\epsilon$. Based on their numerical
simulations and theoretical arguments, \citet{klessen10} argue that
the efficiency factor $\epsilon$ is roughly equal to the ratio between
the density of the accreting gas to the mean density of the bound
system.

Computing the ratio between the CO mass of SGMC~2 and the mass
accretion rate, we obtain a time scale $t_{\rm acc} = 7 \times 10^7
\times f_{\rm H_2}~\mathrm{yr}$.  For $f_{\rm H_2} \sim 0.25$, this is
slightly longer than the 10~Myr time scale over which the dynamics of
the interaction may have been driving the convergent flow
\citep{renaud08}.  However, within the uncertainties in this rough
calculation, we consider that the present accretion rate is close to
the mean rate needed to account for the formation of the SGMC~2
complex as a result of gravitational fragmentation and gas accretion
driven by the galaxy interaction.

Our interpretation introduces a dynamical view of the present state of
the SGMC~2 complex.  The virialized CO complex is physically
associated with gas which is too turbulent to be bound. This unbound
gas is dynamically fed by the convergent flow. It contributes an
accretion flow onto the CO complex, which has the necessary magnitude
to drive the gas turbulence, i.e. to balance the dissipation of the
gas turbulent kinetic energy. \citet{klessen10} have proposed a
similar interpretation to account for the formation and subsequent
growth of turbulent molecular clouds in the LMC. In Sect. \ref{sec:sfe},
we discuss the impact of this interpretation on the star formation
efficiency within SGMC~2.

\subsection{The nature of the compact \hh\ source}
\label{sec:ssc_form}

In this section we extend our interpretation of the extended \hh\
emission to the compact \hh\ source.  Like for the SGMC~2 complex we
propose an interpretation where the \hh\ luminosity traces the
dissipation of the turbulent energy of the gas and gas accretion.

As a first test whether this is a plausible interpretation, we compare
the dissipation rate of the turbulent energy of the gas with the \hh\
luminosity of the source. Combining equations (\ref{eqD}) and
(\ref{eqA}), we obtain an equation that we can use to estimate the
mass of the compact source, $M_{\rm cloud}$, from the \hh\ luminosity.

\begin{equation}
3/2 \, M_{\rm cloud} \times \sigma_{\rm v}^3/R= \epsilon \times L_{\rm H_2} / f_{\rm H_2}  
\end{equation}

\noindent
where $\sigma_{\rm v} = 20 \, {\rm km \, s^{-1}}$ based on the width of the narrow component in the 
CRIRES spectrum ($\rm FWHM = 50 \,  km \, s^{-1}$).  We find 

\begin{equation}
M_{\rm cloud} = 2.7 \times 10^7 \times \epsilon / f_{\rm H_2} \times  (L_{\rm H_2} /2.6 \times 10^6\,L_{\odot})\  M_{\odot}
\end{equation}

This mass estimate depends on the ratio $\epsilon / f_{\rm
H_2}$. Assuming that this ratio is $\sim 1$ as estimated for the CO
complex, we find a mass slightly larger than the virial mass $M_{\rm
vir} = 1.3 \times 10^7\,$\msun\ derived for the same velocity
dispersion in Sect. \ref{sec:cs}. We conclude from this comparison that
the \hh\ emission from the compact source may be accounted for by the
dissipation of turbulent kinetic energy if the source is a
gravitationally bound cloud with a mass in the range 1 to a few
$10^7\,M_\odot $. With this interpretation, the compact \hh\ source is a
gravitationally bound cloud, massive enough to form a super star
cluster.

Like what we have done in the previous section for the CO complex, we
can use equation (\ref{eqA}) to estimate the mass accretion rate and
thereby the cloud formation time scale.  For $\sigma_{\rm H_2}$, we
use the width of the broad component ($\rm FWHM \, = \, 160 \, km \,
s^{-1}$) in the line profile of the 1$-$0 H$_2$ S(1) measured with CRIRES. 
We find:

\begin{equation}
\dot{M}_{\rm acc} =  9.2 \times \left(\frac{f_{\rm H_2}}{0.25}\right)^{-1} \times \left(\frac{L_{\rm H_2}}{2.6 \times 10^6 \, L_{\odot}}\right) \, {\rm M}_{\odot}\ {\rm yr}^{-1}
\end{equation}

The ratio between $M_{\rm cloud}$ and $\dot{M}_{\rm acc} $ yields a
cloud formation time scale by accretion of $t_{\rm acc} = 2.9 \times
10^6 \times (\epsilon/0.25) \,$yr. This value is about twice the cloud
dynamical time $R/\sigma_{\rm v}$. We would thus be observing a
pre-cluster cloud early in its evolution. Even at this early stage it
is remarkable that the fraction of the cloud mass that has been
transformed into stars is very small ($\sim 0.3\%$, Sect. \ref{sec:cs}
). We observe that the cloud has formed and become gravitationally
bound without having formed massive stars.

If our interpretation is correct, by observing the gas cooling through
H$_2$ lines, we may have discovered a massive cloud on its way to form
a SSC within the next few Myr. However, this conclusion is
only tentative, and will remain incomplete, until we obtain the
missing information about the mass, the density structure and the
kinematics of the bulk of the gas. The H$_2$ line emission provides
this information only for the fraction of warm, shock-heated gas. The
missing information can be obtained with ALMA.

\section{Discussion}

Our observations are related to two main questions in the field of
star formation. How do super star clusters form?  What are the roles
of turbulence and stellar feedback in regulating the star formation
efficiency in galaxy mergers?  We discuss how our observations and
interpretation may be of general relevance for these questions. 

\subsection{The formation of super-star clusters}

Star formation is the result of the hierarchical structure of the
molecular interstellar medium established by turbulent compression and
gravitational contraction \citep{maclow04}. The available gas 
fragments over a range of masses and time scales. Massive clusters
form at different times from the densest and most massive gas
concentrations. For SGMC~2, this view is supported by the presence of
two massive clusters (the near-IR SSC and the compact ionized source),
which have formed before the compact \hh\ source.  More clusters, too
small to be identified individually, are likely to have formed or to
be in the process of being formed.  Thus, the formation of super star
clusters in SGMC~2 can be understood within the same framework as the
formation of smaller clusters in disk galaxies like the Milky Way.

Our work highlights three physical parameters which may contribute to
account for the formation of exceptionally massive clusters in the
overlap region of the Antennae. These may be of general relevance for
the formation of super star clusters in other extragalactic objects:
other interacting/merging galaxies, and also starbursts and irregular
dwarf galaxies \citep{keto05,weidner10}.  (1) The presence of
compressive motion on large scaleswhich collects the gas.  The
formation of the SGMC~2 complex and that of the compact \hh\ source is
driven by the galaxy interaction.  The large scale dynamics triggers
the gas compression necessary for their formation and their subsequent
growth by gas accretion.  (2) As discussed in earlier studies
\citep{escala08,weidner10}, shear is the second key parameter in the
formation of massive clusters because it sets the maximum mass that
the gas self-gravity can collect. In the Antennae, the present
geometry of the interaction, which temporarily makes the tidal forces
compressive \citep{renaud08}, is favorable to the formation of massive
clusters.  As discussed by \citet{renaud09} such conditions occur
repeatedly in galaxy mergers.  (3) High turbulence is an additional
key parameter. Since the Jeans mass is proportional to $\sigma^4 $
(see Sect. \ref{sec:gf}), a high value of the velocity dispersion $\sigma$
scales up the masses of the clouds formed by gravitational
fragmentation.  To form a super star cluster, it is not sufficient to
bind a large mass of gas. It is also necessary that the star formation
efficiency becomes high when the gas mass is highly concentrated.  It
is likely that high turbulence is a key factor which prevents star
formation to be efficient before the cloud mass has been concentrated
by gravity. This is a prerequisite to form a dense, potentially bound,
cluster rather than a loose OB association \citep{elmegreen08}.
 
\subsection{The impact of turbulence}

The impact of turbulence on the star formation efficiency is a long
standing topic of research \citep{maclow04}. \citet{krumholz05} and
\citet{padoan11} used numerical simulations to quantify the dependence
of the star-formation rate per free-fall time on the Mach number of
turbulence.  Both agree that, independent of the mean cloud density,
for a cloud near virial equilibrium, the fraction of the gas mass
that is unstable to collapse is small. If this is correct, then the formation of
dense clusters within a cloud supported by turbulence must occur over
several cloud crossing times as argued by \citet{tan06} and
\citet{krumholz06}.  \citet{elmegreen08} expresses a different view
point by proposing that star formation becomes efficient within cores
with a mean density $> 10^4~$H \, cm$^{-3}$ independent of turbulence.

These ideas are being tested against observations of galactic star
forming regions and cores, which are far more detailed than the
present information we have on the compact \hh\ source.  High angular
resolution observations of molecular lines at mm wavelengths, which
unlike the near-IR \hh\ lines will trace the kinematics of the bulk of
the gas and the source density structure, would be needed for a
detailed study of the formation of super star clusters on the example
of this source, and if it is consistent with present ideas based on
observations of lower mass and less turbulent cores.  Here we only
make preliminary remarks that can soon be tested with ALMA.  The mean
density of the compact \hh\ source is $ \sim 10^4~$H \, cm$^{-3}$. We
estimate its formation time scale to at least $10^6 \,$yr, but the
star-formation efficiency is still very small (stellar to gas mass
ratio $\sim 0.3\%$). This remarkable result may indicate that the high
turbulence (1D velocity dispersion $\rm \sim 20 \, km \, s^{-1}$) has
been very effective at preventing star formation.  Star formation may
occur rapidly once the rate of mass accretion becomes insufficient to
drive the present amplitude of turbulence. The fact that stars may be
forming out of gas which has lost the turbulent energy it had during
the initial gravitational contraction reduces the requirement on the
star formation efficiency to form a bound stellar cluster.
 
\subsection{Star formation efficiency}\label{sec:sfe}

The star formation rate (SFR) in the Antennae merger has been
estimated to $\rm 20 \, $\msun \, yr$^{-1}$ for a total molecular gas
mass $\sim 10^{10} \, $\msun. Local values measured from CO and
H$\alpha$ images of the Antennae agree with the general correlation
between the SFR and molecular gas surface densities, the 
Schmidt-Kennicutt law \citep{kennicutt98,zhang01}.  The burst of star
formation observed in the overlap region follows mainly from the high
surface density of molecular gas with only a small (a factor $\sim 3$)
enhancement of the star formation efficiency with respect to the
Galactic value. This enhancement comes from the non-linear dependence
of the SFR on the surface density of gas.  The star formation rate per
unit gas mass in the SINFONI field of view is similar to the global value.
Based on their Spitzer and Herschel data, \citet{brandl09} and
\cite{klaas10} find an SFR of $\sim$~0.7 \msun\ yr$^{-1}$ for a
molecular mass of $4\tten{8}$\msun.

To compute the star formation efficiency, we consider that star
formation is triggered by gas compression and occurs over the
characteristic time, $\sim 10\,$Myr, over which the tidal interaction
is compressive \citep{renaud08}. This is also the crossing time
across the SINFONI field of view for the CO line width. Over this time the
gas mass converted into stars is 7$\tten{6}$~\msun, which yields a star
formation efficiency of 2\%. The SSC accounts for a significant
fraction of this stellar mass (see estimates of the cluster mass in
Table~\ref{tab:sc}). For the SGMC~2 complex, like for the compact
\hh\ source, the low star-formation efficiency may result from the
strong, driven turbulence. For the molecular complex,
the star-formation efficiency may remain low until it is disrupted by
the tidal interaction.  If turbulence continues to be driven by
on-going accretion, it is possible that the CO complex will be
dispersed without losing its turbulent energy. It will be interesting 
to extend the present study to other complexes in the overlap region
to test this idea.

\subsection{Stellar feedback}\label{sec:stef}

It is interesting to compare this analysis and interpretation of
observational data with numerical simulations.  In their Figure~1,
\citet{karl10b} compare the star-formation rates in recent numerical
simulations of the Antennae.  All simulations predict a significant
enhancement of the star-formation efficiency, relative to its value in
the two spirals prior to interaction, after pericenter passage. In the 
simulations, the enhancement in the star formation
efficiency is triggered by the gas compression and subsequent gas
cooling.  For several runs the enhancement is one order of magnitude
or more, i.e. larger than that derived from observations.
\citet{karl10b} present new simulations that quantify the impact of
stellar feedback on the star formation efficiency, and argue that it
is necessary to invoke stellar feedback to compensate for the gas
cooling. Our \hh\ observations do not support this idea.

In Sect. \ref{sec:large}, we argue that the gas turbulence is driven by
the galaxy interaction. Stellar feedback is energetically significant
but not a key contributor to the turbulent energy of the molecular
gas. This holds for the SGMC~2 complex and the compact \hh\ source.
We start discussing feedback on the scale of the SGMC~2 complex using
Starburst99 models. For a Salpeter IMF and continuous star formation
over $10~$Myr, the mechanical energy associated with stellar winds and
supernovae explosions is $L_\mathrm{Mech}(\mathrm{SF}) \sim 4\times
10^{7} \times$ (SFR/0.7 \msun\ yr$^{-1})~L_\odot$. Most of this
energy is released by the stars in the SSC. The non-thermal component of the
radio flux from the SSC \citep{neff00} indicates that the SSC has
evolved to an age where the most massive stars are exploding as
supernovae. For a stellar mass of a few $10^6 \, $\msun, the
mechanical power from stellar winds and supernovae is also a few $\rm
10^7$~$L_\odot$.  This value is one order of magnitude larger than the
\hh\ luminosity of the extended emission, but the lack of enhancement
of the \hh\ line width towards the SSC is evidence that stellar
feedback does not contribute significantly to driving the turbulent
kinetic energy of the \hh\ gas.  The energy from stellar feedback must
be mostly transferred to the hot X-ray emitting plasma
\citep{baldi06}.  We plan to test this tentative conclusion with
additional SINFONI data towards other super star clusters in the
overlap region.

For the compact source, we estimate the star formation rate by
dividing the stellar mass of $4\times 10^4$\msun\ (from
Sect. \ref{sec:cs}) by the formation time scale $t_{\rm acc} \sim 3 \times
10^6\,$yr (from Sect. \ref{sec:ssc_form}).  We find SFR $\sim 10^{-2} \,
M_\odot \, {\rm yr^{-1}}$. For this rate, the mechanical energy from
stellar winds is two orders of magnitude smaller than the \hh\
luminosity of the compact source.  Using formula (6) in
\citet{matzner02} we verified that the contribution from proto-stellar
winds around low mass stars is also much smaller than the \hh\ luminosity.


\section{Conclusions}

We presented an analysis of VLT/SINFONI near-IR imaging spectroscopy
of the region with the brightest \hh\ rotational line emission in the
Antennae overlap region, which has previously been identified with
Spitzer/IRS spectroscopy. The region encompasses one of the supergiant
molecular complexes, SGMC~2, discovered through CO interferometric
observations.  It corresponds to one of the brightest FIR knots in the
overlap region, and is near a young ($\sim$5~Myr) super star cluster
previously identified in NIR imaging including HST imaging. Our
imaging spectroscopy provides us with constraints on the spatial
distribution, kinematics and excitation of the \hh\ and \ion{H}{ii}
gas out to a radius of 300~pc from this cluster. Based on these
observations, we investigated how the large-scale gas dynamics may
trigger the formation of the SSC and regulate the star
formation efficiency. We have also discussed how the observations and
our interpretation may be of general relevance for the formation of
dense massive clusters and the efficiency of star formation in
mergers.  We list our main results. \\

\noindent
$\bullet$
We find extended near-IR \hh\ line emission across much of our SINFONI
field-of-view with broad line widths (${\rm FWHM \sim
200\,km~s^{-1}}$), larger than those measured from CO and \brg\ at the
same position (which are 70 and 130 km s$^{-1}$, respectively). The
line width is commensurate with a large-scale velocity gradient across
the field.  We argue that this extended emission component traces a
convergent turbulent flow driven by the galaxies interaction. Spectral
diagnostics show that the \hh\ emission is shock powered and traces
the dissipation of the gas turbulent kinetic energy. \\

\noindent
$\bullet$
The data reveal a compact \hh\ source (50~pc diameter) with a $K$-band
spectrum showing only \hh\ line emission.  The \hh\ lines are
spectrally resolved with a width $\sim 150\,$ km~s$^{-1}$ (FWHM).  The
\hh\ emission from this source is also shock excited.  The cloud
virial mass is $\sim 1\tten{7} \, $\msun.  The absence of \brg\
emission and of some obvious counterpart in the radio continuum set a
low limit on the mass of newly formed stars (stellar mass fraction
$\sim 0.3\%$).  To our knowledge, this is the first time that an
extragalactic source with such characteristics is identified.\\

\noindent
$\bullet$
The width of the \hh\ spectra show that the SGMC~2 complex and the
compact source are both associated with gas which is too turbulent to
be bound. We argue that this suggests that the \hh\
emission is powered by gas accretion. We show that the required
accretion rate is of the right order of magnitude to drive the gas
turbulence and to account for the formation of both the SGMC~2 complex
and the compact source by accretion.\\

\noindent
$\bullet$
By observing gas cooling through \hh\ lines, we may have discovered a
massive cloud on its way to form an SSC within the next few Myr.
However, this conclusion is only tentative, and will remain
incomplete, until we obtain the missing information about the mass,
the density structure and the kinematics of the bulk of the gas. The
\hh\ line emission provides this information only for the warm shock
excited fraction.  The missing information can be obtained with ALMA.\\

\noindent
$\bullet$
We propose that the strong turbulence observed in the extended
emission and the compact \hh\ source is of general relevance for the
formation of SSCs in two ways. A high value of the gas velocity
dispersion increases the masses of the clouds formed by gravitational
fragmentation. It may also prevent star formation to be efficient
before the cloud is fully formed, i.e. as long as turbulence is driven
by accretion. Clusters may form out of gas which has lost much of the
turbulent energy it had during gravitational formation of the
pre-stellar cloud. If this is correct, then it reduces the requirement
on the star-formation efficiency to form a bound stellar cluster. \\

\noindent
$\bullet$
Our observations highlight the role of merger-driven turbulence in
regulating the star-formation efficiency in interacting
galaxies. Within the field of view of our SINFONI observations, about
2\% of the molecular gas has been turned into stars. We relate this
small efficiency to the fact that the gas turbulence does not
dissipate because it continues to be driven by on-going accretion.

%
\begin{acknowledgements}
The authors wish to thank the staff at the VLT and at the CFHT for
making these observations. We are grateful to C. Wilson for providing
us with her OVRO CO data. We would like to thank the anonymous referee
for the detailed and constructive report that helped us in the
interpretation of our data. We thank B. Elmegreen, P. Guillard,
P. Hennebelle and G. Pineau des F\^orets for helpful discussions at an
early stage in the writing of this paper.  C.H. acknowledges support
from a CNRS-CONICYT scholarship.  This research is funded by CONICYT
and CNRS, in accordance with the agreement written on December 11,
2007.
\end{acknowledgements}
\bibliographystyle{aa}
\bibliography{papers}
\end{document}